%Paper: gr-qc/9304044
%From: Ranjeet S. Tate <rstate@cosmic.physics.ucsb.edu>
%Date: Thu, 29 Apr 93 17:44:35 -0700

%Plain TeX file, 7 pages

\def\0{\emptyset}

\def\M{{\cal M}}
\def\L{{\cal L}}
\def\A{{\cal A}}
\def\G{{\cal G}}

\def\S{\Sigma}
\def\l{\lambda}
\def\a{\alpha}
\def\b{\beta}

\def\Tr{\rm Tr}
\magnification = \magstep1

\centerline{\bf Completeness of Wilson loop functionals}
\centerline{\bf on the moduli space of $SL(2,C)$ and $SU(1,1)$-connections}
\centerline{gr-qc/9304044}
\bigskip
\centerline{Abhay Ashtekar${}^1$ and Jerzy Lewandowski${}^{1,2}$}
{\it \centerline{${}^1$ Department of Physics, Syracuse University,
Syracuse, N.Y. 13244-1130}}
{\it \centerline{${}^2$ Department of Physics, University
of Florida, Gainsville, FL 32611}}
\bigskip
{\sl Abstract.}

The structure of the moduli spaces $\M := \A/\G$ of (all, not just
flat) $SL(2,C)$ and $SU(1,1)$ connections on a n-manifold is analysed.
For any topology on the corresponding spaces $\A$ of all connections
which satisfies the weak requirement of compatibility with the affine
structure of $\A$, the moduli space $\M$ is shown to be non-Hausdorff.
It is then shown that the Wilson loop functionals --i.e., the traces
of holonomies of connections around closed loops-- are complete in the
sense that they suffice to separate all separable points of $\M$. The
methods are general enough to allow the underlying n-manifold to be
topologically non-trivial and for connections to be defined on
non-trivial bundles. The results have implications for canonical
quantum general relativity in 4 and 3 dimensions.
\bigskip
\medskip

\noindent{\sl 1. Introduction}

The structure of the moduli spaces $\M := \A/\G$ of connections has been
studied in detail in the case when the gauge group $G$ is compact and has
been shown to admit the structure of an infinite dimensional manifold
except for ``conical singularities'' at those points where the connections
admit symmetries (so that the holonomy group is a proper sub-group of the
full gauge group).
\footnote{${}^1$}{This structure is analogous to that of Wheeler's superspace
of 3-geometries, which had been analysed by Fisher, Marsden and others
already in the seventies.}
(See, e.g. [1].) In the non-compact case, on the other hand, relatively
little seems to be known. From a physical standpoint, this was not considered
to be handicap because one can restrict oneself to the compact case
in realistic gauge theories. In recent years, however, general relativity
in 3 and 4 dimensions has been recast as a theory of connections (see, e.g.,
[2,3]), and the relevant gauge groups -- $SU(1,1)$ and $SL(2,C)$ respectively--
are non-compact. It is therefore of considerable physical interest to extend
the previous work and analyse the structure of the moduli spaces of
corresponding connections.

The issue of completeness of the Wilson loop functionals was analysed in
detail recently [4]. While for $SU(2)$-connections, these functionals separate
all points of $\M$, for $SL(2,C)$ and $SU(1,1)$-connections, this is not the
case; the Wilson loop functionals now fail to capture the full gauge invariant
information in the connections. This failure can occur when the connection
is reducible, i.e. only on ``sets of measure zero'' in $\M$. Nonetheless,
this limitation is significant in quantization of the theory since the
``missing information'' can lead to physically irrelevant superselection rules
[3,5].

In this Letter, we will show that the failure occurs simply because the
points in question are not separable in any reasonable topology. Thus, the
Wilson loop functionals are in fact ``as complete as they can be.'' The
implications of this result to the quantization procedure are not yet fully
understood because we have very little experience in quantizing systems whose
configuration spaces fail to be Hausdorff. On the mathematical side, on the
other hand, the ramifications of these results seem more transparent.
Since non-Hausdorffness occurs at certain reducible connections, it is
tempting to conjecture that in the passage from compact gauge groups to
non-compact, extra care would be needed only at such connections. While in
the compact case $\M$ fails to have a nice differential structure at these
points, in the non-compact case, problems may arise already at the topological
level. In the compact case, the failure occurs because the orbits in $\A$ of
the gauge group through these connections are ``thinner'' than generic orbits.
In the non-compact case, not only are they thinner but they may even be
contained in the closure of other orbits.

In application to 4 (and 3)-dimensional general relativity, the $SL(2,C)$
(respectively $SU(1,1)$) connections are defined on 3 (respectively
2)-dimensional manifolds, the Cauchy surfaces. In this Letter, however, we
will consider the general case and consider connections on any principal
$SL(2,C)$ or $SU(1,1)$ bundle over an n-dimensional real manifold $\S$. We
will begin with some preliminaries, then explain the origin of the
non-Hausdorff character using a trivial bundle and finally establish the main
theorem in full generality.
\bigskip

\noindent {\sl 2. Preliminaries}

Standard definitions and statements about bundles and connections are
available from Kobayashi and  Nomizu [6] and  Steenrod [7]. We denote by
$\A$ the set of connections defined on a principal fibre bundle $P(\S,G)$
with the structure group $G$ which is either $SL(2,C)$ or $SU(1,1)$.
Following the notation introduced in [8], which has become standard in
quantum general relativity, we will denote the Wilson loop functional
associated with a closed loop $\alpha$ by $T_\alpha$. Thus, associated with
a piecewise $C^1$ loop
$$\a: [0,1]\rightarrow \S,$$
with $\a (0) = \a(1)$, we have a function on $\A$:
$$T_\a(A) = \textstyle{1\over 2} TrH(\a,A)$$
where $H(\a,A)$ is an element of $G$ assigned to $\a$ and $A$ by the holonomy
map. (Although $H(\a,A)$ depends on the choice of a point in the fiber of
$P(\S, G)$ over $\a(0)$, the $ TrH(\a,A)$ is uniquely defined). Since $T_\a$
is invariant with respect to the group $\G$ of gauge transformations acting
on $\A$, we can consider it as a function on the quotient $\M := \A/\G$.

We can now specify our topological assumption. We assume that $\A $ is
equipped with a topology compatible with the affine structure defined
on the  space of connections; i.e. that every line in $A $,
$$A(t)= tA_1 + (1-t)A_2, \ \ A_1,\ A_2\in A ,\eqno(1)$$
is continuous. This is a very weak assumption. In practice, one normally
equips $\A$ with the structure of a suitable Sobolev space [1] and then our
assumption is trivially satisfied. The topology on $\M$ is induced
by this topology on $\A$ via the quotient construction.

The origin of the non-Hausdorff character of $\M$ can be seen rather easily in
the case when the bundle is trivial. Let $(\tau_1, \tau_2, \tau_3)$  be a
basis in $su(2)$ which is orthonormal with respect to the scalar product
given by $-{1 \over 2}\Tr$. (Thus, the $\tau_i$ are $i$ times the Pauli
matrices). Next, define null basis:
$$\tau_+ := \tau_1 + i\tau_2,\ \quad  \tau_- := \tau_1 - i\tau_2. $$
We consider hereafter $sl(2,C)$ as a complexification of $su(2)$ and $su(1,1)$
as a real sub-algebra of $sl(2,C)$ generated by $(\tau_+, \tau_-, i \tau_3)$,
and extend this identification to the level of groups. Consider a connection
which (when pulled down by some global section) is given by the following (Lie
algebra)-valued 1-form
$$A = A^+\tau_+ + A^3\tau_3\eqno(2)$$
$A^+$ and $A^3$ being arbitrary complex 1-forms on $\S$.
The gauge orbit passing through $A$ includes a line
$$A (\l )= e^{-2\l}A^+\tau_+ + A^3\tau_3, $$
which is the image of $A$ under the action of the 1-dimensional subgroup of
$SU(1,1)$, represented in this gauge by the constant $SU(1,1)$-valued functions
 $$g_{\l}:= e^{i\l\tau_3},\eqno(3)$$
where the real $\l$ is a parameter in the subgroup. But in the limit, we
have:
$$\l\rightarrow \infty,\ \ \  A(\l)\rightarrow A^3 \tau_3.$$
It therefore follows that for every continuous and gauge invariant function
$f$ on $\A $, we must have:
$$f(A^+\tau_+ + A^3 \tau_3) = f( A^3 \tau_3).\eqno(4)$$
Note that the connections $A^+\tau_+ + A^3 \tau_3$ and $A^3 \tau_3$ have
distinct holonomy groups and therefore define distinct points of $\M$.
Eq.(4) implies that these points can not be separated; $\M$ is not
Hausdorff. This, incidentally, can be regarded as a ``topological
explanation'' of the fact that the loop variables $T_{\a}$ in quantum
general relativity are insensitive to the term proportional to $\tau_+$ if
the connection has the form (2).
\bigskip

{\sl 3. Main Result}

Our aim now is to show that  the set of all the functions $T_\a$ separates
all the {\it separable} points of $\M$. Let us begin by fixing the notation.
Denote by $L$ the set of piecewise $C^1$ loops in $M$. Next, given a connection
$A\in\A $ we will denote its holonomy group by $G_H(A)$ and define its
degeneracy, Deg$(A)$, as follows:
$${\rm Deg}(A) := \{A'\in\A |\ {\rm for}\ {\rm every}\ \a\in \L , \ \ T_\a(A')
= T_\a(A)\}. $$
We will let $A\G$ stand for the orbit in $\A$ of the (local) gauge group $\G$
which contains $A$. Note that, since every $T_\a$ is a gauge invariant
function on $\A$, ${\rm Deg}(A)$ contains the entire orbit $A\G$. Finally,
two sub-groups of $SL(2,C)$ (respectively $SU(1,1)$) will play an important
role in what follows. First is the {\it group of null rotations} to be denoted
by $G(+,3)$. This is the group generated by the Lie algebra of complex
(respectively, real) linear combinations of $(\tau_+, \tau_3)$. Similarly, we
will denote by $G(+)$ the group generated by the Lie algebra of complex (real)
multiples of $\tau_+$ and by $G(3)$ the group generated by the Lie algebra of
complex (real) multiples of $\tau_3$.

\bigskip\goodbreak
The main result can be stated as follows:

\noindent
{\bf Theorem } Suppose that $A_1, A_2\in \A $ and
  $$T_\a(A_1)=T_\a(A_2) $$
for every loop $\a \in L$. Then, for every continuous and gauge invariant
function $f$ defined on $\A$, we have:
  $$f(A_1) = f(A_2).$$
\medskip
{\it Proof:}
The proof consists of three steps which we extract in the form of lemmas
stated below. The key issue is: i) whether there exist connections $A$ for
which $A\G$ is smaller than ${\rm Deg}(A)$; and, if this happens, ii) whether
the point of $\M$ defined by $A$ is non-Hausdorff, i.e., whether the
closure $\overline{A\G}$ of $A\G$ contains other gauge orbits $A_0\G$.
\medskip
\noindent {\bf Lemma 1}  The property $A\G < {\rm Deg}(A)$ holds if and only
if the holonomy group $G_H(A)$ of $A$ is a subgroup of the group of
null rotations $G(+,3)$.
\medskip
\noindent {\bf Lemma 2} If the holonomy group $G_H(A)$ of $A$ is a subgroup
of $G(+,3)$, then there exists a unique gauge orbit $A_0\G \subset
{\rm Deg}(A)$ such that $G_H(A_0)\subset G(3)$.
\medskip
\noindent {\bf Lemma 3} Suppose that the holonomy group $G_H(A)$
of a connection $A\in \A$ is a subgroup of $G(+,3)$. Then, in the closure
${\overline{A\G}}$ of the orbit $A\G$, there is a connection
$A_0$ such that $G_H(A_0)\subset G(3)$ and $A_0\in {\rm Deg}(A)$.

It follows from the above lemmas that if $T_\a$ fail to separate a point
of $\M$, i.e., if there exists $A \in \A$ such that $A\G <{\rm Deg}(A)$, then
there is a unique gauge orbit $A_0\G$ in Deg$(A)$ which is contained
in the closure $\overline{A\G}$ of $A\G$. Therefore, for any $A_1,\ A_2 \in
{\rm Deg}(A)$, we have:
  $${\overline {A_1\G}}\cap {\overline {A_2\G}} \not=\0 ,$$
because the intersection contains the connection $A_0$. Since a gauge
invariant and continuous function $f$ on $\A $ is constant on the
closed of orbits, it is necessarily true that $f(A_1) = f(A_2).\qquad
\qquad\bullet$

\medskip
{\it proof of Lemma 1:}
The analysis of the invertibility of the mapping
  $$H(.,A)\rightarrow T_.(A) $$
for a connection $A$ which has a connected holonomy group has been performed
in [4]. It was shown there that, unless $G_H(A)\subset G(+,3)$, we can
reconstruct the element $H(\a,A)$ of $G$ provided that we know the value
$T_\b(A)$ for every loop $\b\in L$. Thus, to establish the Lemma, we need
only consider the disconnected subgroups of $SL(2,C)$ that can arise as
holonomy groups. These were classified by Jacobson and Romano [9]. The only
subgroup of $SL(2,C)$ which is not contained in $G(+,3)$ is that denoted
in [9] by $G(3,Z_2)$. This is the union of two connected components: $G(3)$
and $G(3)\circ\tau_2$ where $\tau_2$ is now regarded as an element of
$SL(2,C)$. But if the holonomy mapping takes values in this group then there
exists a loop $\a_1$ such that
  $$T_{\a_1} = 0,\quad  T_{\a_1\circ\a_1} = -1$$
(Actually, the first equality above implies the second one.)
Thus we can identify the holonomy group. But then we know that in some gauge,
every $H(\a,A)$ is either diagonal or antidiagonal. Moreover, modulo $G(3)$
gauge transformations,  $H(\a_1,A)$ is just $\tau_2$. Finally, from values of
$T_.(A)$ taken on suitable products of loops, we can easily recover
$H(\a,A)$, whence, for such a connection $A$, Deg$(A) = A\G$.
\medskip

{\it Proof of Lemma 2:}
Suppose that the holonomy group of a connection $A'$ is a subgroup of $G(+,3)$.
Then we can find another $A$ gauge equivalent to $A'$ such that the holonomy
map of $A$ takes values in $G(+,3)$ and  has the form
$$H_\a(A) = {\rm cos}\theta_\a(A) + \tau_3 {\rm sin}\theta_\a(A) +\tau_+
\phi_\a(A)$$
where $\theta_\a$ and $\phi_\a$ are  complex-valued functions of $A$
($\theta_\a$ not necessarily continuous). We define a map
$$\tilde{H} : L\ni \a \rightarrow H_\a(A) - \tau_+ \phi_\a(A)\in  G(3).
\eqno(5)$$
It not difficult to check that $\tilde{H}$ satisfies all the conditions [10]
sufficient for the existence of a connection $A_0$ such that
$\tilde H (\a)$ coincides with the holonomy mapping  $H(\a,\ A_0)$.
Furthermore, $A_0\in {\rm Deg}(A)$, since by (5) for every loop $\a$
$$T_\a(A_0)= T_\a(A).\eqno(6)$$
This establishes the existence. The uniqueness of a $G(3)$ connection
satisfying (6) follows from the fact that, up to  gauge transformations,
$A_0$ can be completely reconstructed from $T_\a$'s.
\medskip

{\it Proof of Lemma 3:} The idea of the proof is to find a one parameter
subgroup of gauge transformations analogous to (3), allowing, however, for
the bundle to be non-trivial. Now, there exists an open covering $\{V_I\}$
on $\S$ and local sections
$$s_I: V_I\rightarrow P, $$
such that
$$A_I := s_I*A = A_I^3\tau_3 + A_I^+\tau_+,\eqno(7)$$
which means that locally defined $A_I$'s take values in the Lie algebra of
$G(+,3)$. Moreover, every $G(+,3)$ principal bundle over $\S$ is reducible to
a $U(1)$ principal bundle because, topologically, $G(+,3)/U(1)\equiv R^3$
(Rendall [11], Steenrod [7]). Therefore, we can choose the sections $s_I$
in such a way that the transition functions $a_{IJ}$, given by $s_Ia_{IJ}=
s_J $ take values in $U(1)$. Therefore, the part of $A$ in (7) proportional to
$\tau_3$ itself defines a connection $A_0$ on $P$, s.t.
$$s_I^*A_0:= A_I^3\tau_3.  \eqno(8)$$
We can now find a 1-parameter family of automorphisms on the bundle $P$
which, in the limit as the parameter tends to infinity, squeezes $A$ to $A_0$.
Let
$$\psi_\l(x) := e^{i\l\tau_3}$$
where $\l$ is a real constant. By using the sections $s_I$ we lift $\psi_\l$
to a well defined constant function on the holonomy bundle of $A$. Next, we
determine $\psi_\l$ at any point of $P$ by the condition that $\psi_\l(pg) =
g^{-1}\psi(p)g.$ Hence, $\psi_\l$ defines an automorphism of $P$. In addition,
applying $\psi_\l$ to $A$ we obtain
$$\psi_\l^*A =A_0 +  e^{-2\l} (A-A_0). $$
By taking the limit $\l\rightarrow\infty$ we see that
$$A_0\in {\overline {A\G}}. $$
On the other hand, we see from (7) and (8) and from the fact that the
transition
functions are $U(1)$ valued that $T_\a(A_0) = T_\a(A)$ for any loop $\a$. Thus,
we have:
$$A_0\in Deg(A) \quad {\rm and} \quad G_H(A_0)\subset G(3), $$
(whence $A_0\G$ is the unique gauge orbit of Lemma 2). This completes the
proof of Lemma 3 and hence of the Theorem.
\bigskip
{\it Remarks:}
\item{1.} Note that, in the above analysis, we have {\it not} assumed that
the Wilson loop functionals $T_\a$ are continuous on $\M$. If they are --as
is the case if one uses a standard topology [1] on $\A$-- the Theorem has a
stronger implication: $\M$ is non-Hausdorff {\it only} at those points which
can not be separated by the $T_\a$. Furthermore, the arguments used in the
proof provide a classification of these points. We have a natural projection
$G(+,3)\rightarrow G(3)$ and the $G(3)$ part of $H(A,.)$ coincides with
$H(A_0,.)$ which in turn characterizes Deg$(A)$.
\item{2.} It is important to note the sense in which the Wilson loop
functionals have been shown to be complete: they suffice to separate all
separable points of $\M$. In the physics literature, one often assumes
completeness in a different sense, namely that ``all (relevant) gauge
invariant functionals of connections can be expressed as a limit of
polynomials of the Wilson loop functionals.'' While for finite
dimensional manifolds, the two senses of completeness are essentially
equivalent, in the case of $\M$, we do {\it not} have a corresponding result.
\item{3.} In 4-dimensional general relativity, one can associate 4-metrics
with  points of $\M$. Remarkably, the time  evolution given by Einstein's
equations preserves the holonomy group and hence, in particular, the
degeneracy of a point in $\M$. (See [4] for the treatment of the vacuum case
and [9] for the case with a non-vanishing cosmological constant and
topological nontrivialities.) The non-separable points of $\M$ correspond
to  metrics which admit a covariantly constant spinor direction [4]. The
Einstein Equations in this class of metrics has been solved completely
(see [12] for the vacuum case and [13] for the case with a cosmological
constant). These metrics are, in a certain sense, the $(-+++)$ analogs of
the K\"ahler metrics with Euclidean signature [14].

\bigskip\vfill\break
{\sl Acknowledgements.}
This work was supported in part by the National Science Foundation contracts
PHY 86-12424 and PHY 9107007, by the Polish Committee for Scientific Research
(KBN) through grant 2 0430 9101, and, by research funds provided by Syracuse
University.

\bigskip
{\sl References.}
\medskip
\item{[1]} Mitter P K and Viallet C M 1981 {\it Commun. Math. Phys.}
{\bf 79} 43
\item{[2]} Ashtekar A 1987 {\it Phys. Rev. D} {\bf 36} 1587; Ashtekar A
Husain V  Rovelli C Samuel J and Smolin L 1989 {\it Class. Quantum  Grav.}
{\bf 6} L185; Capovilla R  Dell J and Jacobson T 1991 {\it Class. Quantum
Grav.} {\bf 8} 59
\item{[3]} Ashtekar A  1991  {\it Non-perturbative Canonical Quantum Gravity}
(Notes prepared in collaboration with R.S. Tate) (Singapore: World Scientific)
\item{[4]} Goldberg J N  Lewandowski J and Stornaiolo C  1992
{\it Commun. Math. Phys.} {\bf 148}, 377
\item{[5]} Ashtekar A  Tate R and Uggla C 1993 {\it Int. J. Mod. Phys.} {\bf D}
(in press)
\item{[6]} Kobayashi S and Nomizu K 1963  {\it Foundations of Differential
Geometry, Vol 1} (New York: Interscience)
\item{[7]} Steenrod N E 1951 {\it The Topology of Fibre Bundles} (Princeton:
Princeton University Press)
\item{[8]} Rovelli C  and Smolin L 1990 {\it Nucl. Phys.} {\bf B331} 80
\item{[9]} Jacobson T and Romano J D 1992 pre-print UMDGR-92-208
\item{[10]} Lewandowski, J 1993  {\it Class. Quantum Grav.} {\bf 10} 1
\item{[11]} Rendall A 1992 (private communication)
\item{[12]}Goldberg, J and  Kerr R P 1961 {\it J. Math. Phys.} {\bf 2} 327
\item{[13]} Lewandowski J. 1992  {\it Class. Quantum Grav.} {\bf 9} L147
\item{[14]} Nurowski, P 1992  (private communication)

\bye